\documentclass[twocolumn,conference]{IEEEtran}
\usepackage[T1]{fontenc}
\usepackage[latin9]{inputenc}
\usepackage{units}
\usepackage{url}
\usepackage{amstext}
\usepackage{graphicx}
\PassOptionsToPackage{normalem}{ulem}
\usepackage{ulem}
\usepackage[unicode=true,
 bookmarks=true,bookmarksnumbered=true,bookmarksopen=true,bookmarksopenlevel=1,
 breaklinks=false,pdfborder={0 0 0},pdfborderstyle={},backref=false,colorlinks=false]
 {hyperref}
\hypersetup{pdftitle={Your Title},
 pdfauthor={Your Name},
 pdfpagelayout=OneColumn, pdfnewwindow=true, pdfstartview=XYZ, plainpages=false}

\makeatletter
\usepackage[caption=false,font=footnotesize]{subfig}
\usepackage{lipsum}

\makeatother

\begin{document}

\title{Using any Surface to Realize a New Paradigm for Wireless Communications }

\author{Christos Liaskos\IEEEauthorrefmark{1}, Ageliki Tsioliaridou\IEEEauthorrefmark{1},
Andreas Pitsillides\IEEEauthorrefmark{3}, Sotiris Ioannidis\IEEEauthorrefmark{1},
and Ian Akyildiz\IEEEauthorrefmark{2}\IEEEauthorrefmark{3}\\
{\small{}\IEEEauthorrefmark{1}Foundation for Research and Technology
- Hellas (FORTH)}\\
{\small{}Emails: \{cliaskos,atsiolia,sotiris\}@ics.forth.gr}\\
{\small{}\IEEEauthorrefmark{2}Georgia Institute of Technology, School
of Electrical and Computer Engineering}\\
{\small{}Email: ian@ece.gatech.edu}\\
{\small{}\IEEEauthorrefmark{3}University of Cyprus, Computer Science
Department}\\
{\small{}Email: Andreas.Pitsillides@ucy.ac.cy}}
\maketitle

\section{Introduction\label{sec:Introduction}}

Wireless communications have undeniably shaped our ever-day lives.
We expect ubiquitous connectivity to the Internet, with increasing
demands for higher data rates and low lag everywhere: at work, at
home, on the road, even with massive crowds of fellow Internet users
around us. Despite impressive breakthroughs in almost every part of
our wireless devices-from antennas and hardware to operating software-this
demand is getting increasingly challenging to address. The humongous
scale of research effort and investment in the upcoming 5$^{\text{th}}$
generation of wireless communications (5G) reflects the scale of the
challenge~\cite{Akyildiz.2016}. A valuable and unnoticed resource
could be exploited to meet this goal.

A common denominator in related research efforts is that the wireless
environment, i.e., the set of physical objects that stand between
two wireless communicating devices, remains a passive spectator in
the data exchange process. The ensuing effects on the data communication
quality are generally degenerative: First, a transmitting device emits
electromagnetic energy\textendash carrying encoded information\textendash which
dissipates astoundingly fast within the environment. This \emph{path
loss} phenomenon can be envisioned as distributing the same power
over an ever-growing sphere. The power of the intended signal quickly
diminishes, making its reception progressively harder. Second, as
this ever-growing sphere reaches objects, such as walls, doors, desks,
and humans, it scatters uncontrollably in multiple directions. This
creates the \emph{multi-path} phenomenon where many, unsynchronized
echoes of the original signal reach the receiver at the same time,
making it difficult to discern the original. Third, the scattered
signals naturally reach unintended recipients, increasing their noise
levels (and allowing for eavesdropping). Finally, mobile wireless
devices acquire a false perception of the frequency of electromagnetic
waves, a phenomenon known as Doppler effect. Notice that the hunt
for higher data rates in 5G pushes for very high communication frequencies,
e.g., at $60\,\text{GHz}$, where the described effects become extremely
acute~\cite{Akyildiz.2016}.

This article introduces an approach that could tame and control these
effects, producing wireless environment with software-defined electromagnetic
behavior. We investigate the novel idea of \emph{HyperSurfaces}, which
are software-controlled metamaterials embedded in any surface in the
environment~\cite{Liaskos.2015b}. In simpler words, HyperSurfaces
are materials that interact with electromagnetic waves in a fully
software-defined fashion, even \emph{unnaturally}~\cite{Chen.2016,Lim.2016}.
Coating walls, doors, furniture and other objects with HyperSurfaces
constitutes the overall behavior of an indoor wireless environment
\emph{programmable}. Thus, the electromagnetic behavior of the environment
as a whole can become deterministic, controlled and tailored to the
needs of mobile devices within it. The same principle is also applicable
to outdoor settings, by exemplary coating polls or building facades.

The concept of programmable wireless environments can impact wireless
communications immensely, by mitigating\textendash and even negating\textendash path
loss, multi-path and interference effects. This can translate to substantial
gains in communication quality, communication distance and battery
savings of mobile devices, and even in security and privacy. Furthermore,
due to the underlying physics, HyperSurfaces have no restriction in
terms of operating communication frequency, which can extend up to
the terahertz ($\text{THz}$) band~\cite{Tassin.2013}. This attractive
trait constitutes them applicable to a number of cutting edge applications,
such as 4G and 5G, Internet of Things (IoT) and Device-to-Device systems.

\section{The programmable wireless environment concept}

\begin{figure*}[t]
\begin{centering}
\includegraphics[clip,width=1\textwidth]{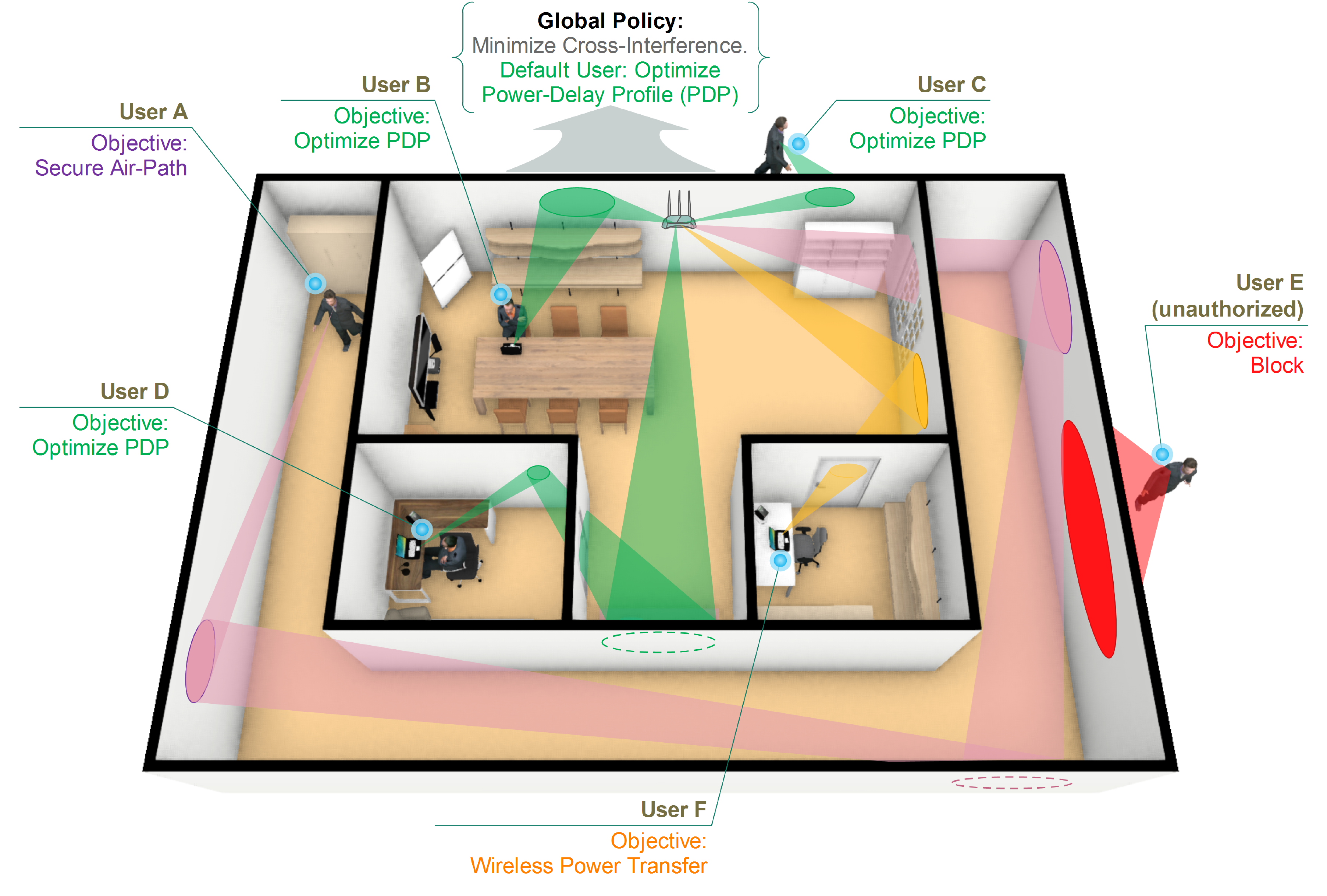}
\par\end{centering}
\caption{\label{fig:example}Illustration of the programmable wireless environment
concept. The electromagnetic behavior of walls is programmatically
changed to exemplary maximize data rates (green use-cases), wireless
power transfer (orange use-case), negate eavesdropping (purple use-case)
and provide electromagnetic shielding (red use-case).}
\end{figure*}

Consider an ever-day communication scenario with multiple users within
a physical environment, as shown in Fig.~\ref{fig:example}. The
common, non-programmable environment is oblivious to the user presence
and their communication needs. The electromagnetic energy simply dissipates
throughout the space uncontrollably, attenuating really fast, causing
interference among devices and allowing for eavesdropping.

In the novel, programmable environments, HyperSurface-coated walls
and objects become connected to the Internet of Things. As such, they
can receive software commands and change their interaction with electromagnetic
waves, serving the user needs in unprecedented ways. In the example
of Fig.~\ref{fig:example}, user A expresses a need for security
against eavesdropping. The programmable environment, in collaboration
with the user devices, sets an improbable ``air-path'' that avoids
all other users, hindering eavesdropping. Users B, C and D express
no requirement, and are automatically treated by a global environment
policy instead, which dictates the optimization of their data transfer
rates. This can be attained by negating cross-interference and a minute
crafting of the received power delay profile (PDP), i.e., ensuring
that all received wave echoes get constructively superposed at the
devices. User F is observed to be inactive and\textendash according
to his preferences\textendash has his device remotely charged by receiving
a very focused energy beam. Finally, user E fails to pass the network's
access policies (e.g., unauthorized physical device address), and
is blocked by the environment. This can be accomplished by absorbing
his emissions, potentially using the harvested energy to a constructive
use.
\begin{figure}[t]
\begin{centering}
\includegraphics[width=1\columnwidth]{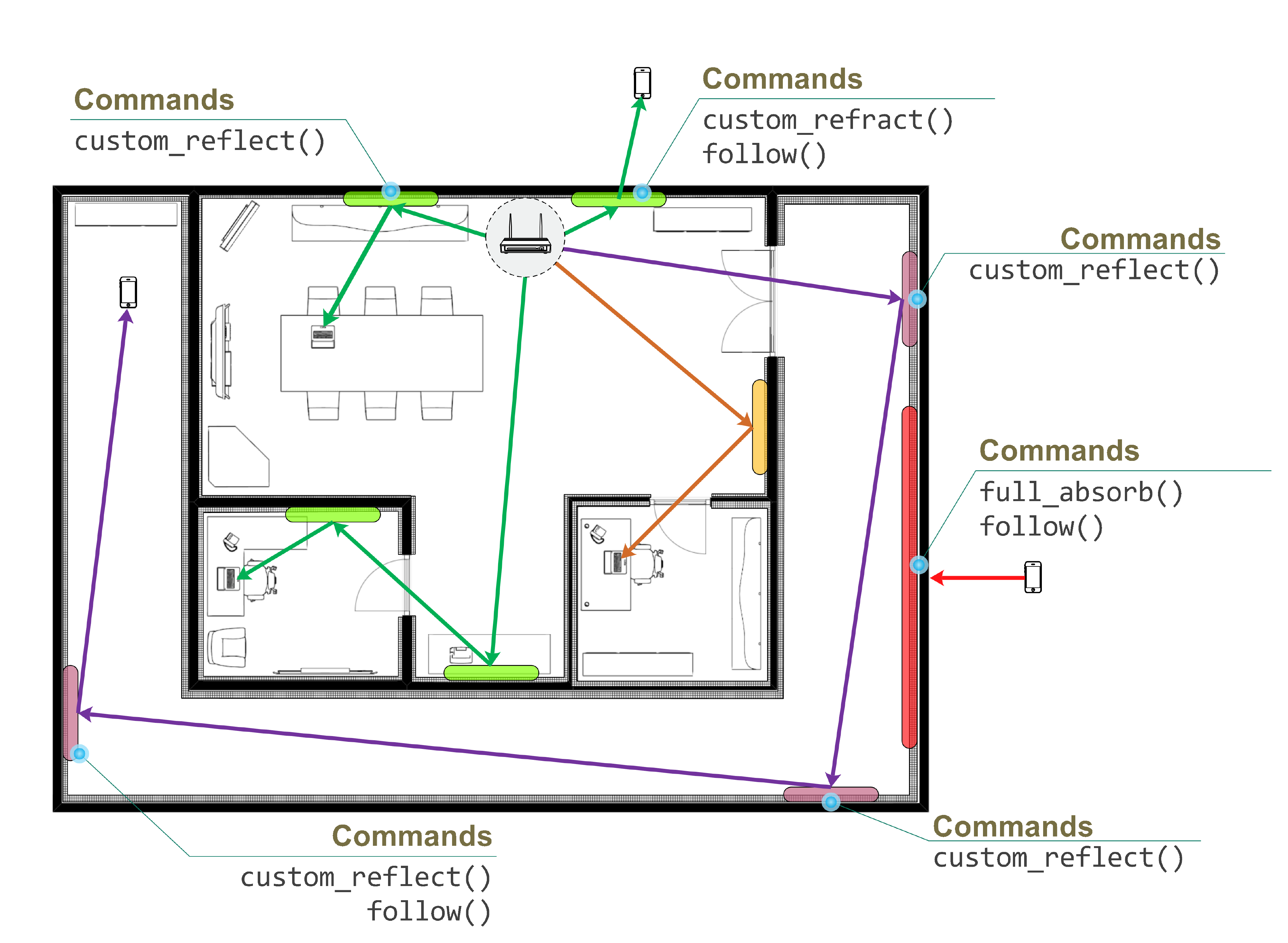}
\par\end{centering}
\caption{\label{fig:2dfloorplan}Software commands are combined and applied
locally on walls, to achieve the objectives of Fig. \ref{fig:example}.}
\end{figure}

From another innovative aspect, the programmable environment concept
abstracts the underlying physics of wireless propagation, exposing
a software programming interface to control it instead. Thus, the
physics behind wireless propagation are brought into the realm of
software developers, treating the electromagnetic behavior of objects
with simple commands, as shown in Fig.~\ref{fig:2dfloorplan}.

Essentially, the HyperSurface-coated objects are treated as ``routers'',
that can forward or block electromagnetic waves in fashion that is
very similar to the concept of routers and firewalls in wired networks.
Connecting devices becomes a problem of finding a route connecting
HyperSurfaces, subject to performance requirements and user access
policies.

\section{Structuring programmable wireless environments}

\begin{figure*}[!t]
\begin{centering}
\includegraphics[clip,width=1\textwidth]{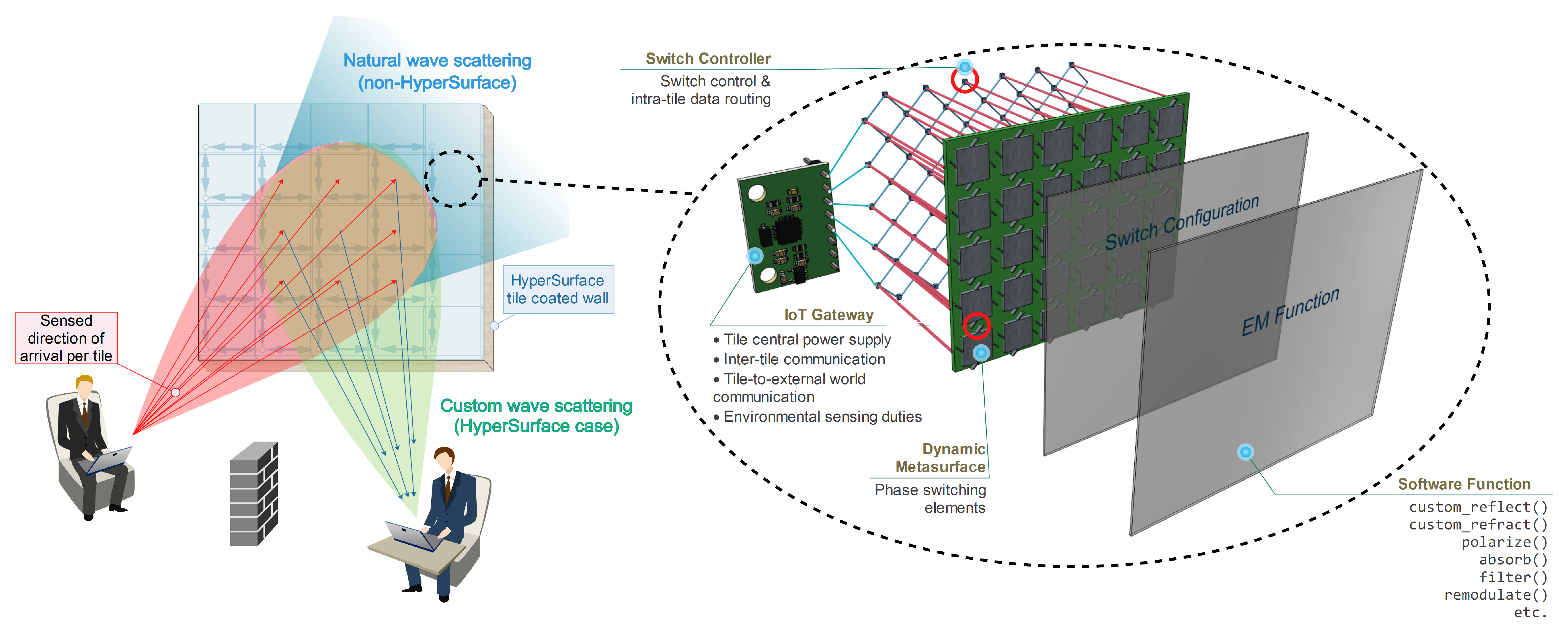}
\par\end{centering}
\caption{\label{fig:workflow}The proposed workflow involving HyperSurface
tile-coated environmental objects. The wireless propagation is tailored
to the needs of the communication link under optimization. Unnatural
propagation, such as lens-like focus and negative reflection angles
can be employed to mitigate path loss and multi-path phenomena, especially
in challenging non-line-of-sight cases. The main tile components are
shown to the right.}
\end{figure*}
The main components that comprise a single HyperSurface tile and imbue
it with control over wave propagation are shown in the right inset
of Fig.~\ref{fig:workflow}. Dynamic metasurfaces are the core technology
for introducing programmable wireless environments. They constitute
the outcome of a research direction in physics interested in creating
materials with engineered electromagnetic properties. Most commonly,
they comprise a metallic pattern, called \emph{meta-atom}, periodically
repeated over a dielectric substrate, and connected via switching
elements~\cite{Li.2016}. The macroscopic electromagnetic interaction
of a metasurface is fully defined by the form of the meta-atoms and
the state of the switches. A certain state of switches may correspond
to full absorption of all impinging waves from a given direction of
arrival, while another may fully reflect them at an unnatural, completely
custom angle~\cite{Chen.2016}.

The translation of switch states to interaction types is performed
by a novel software class, the electromagnetic \emph{compiler}. The
compiler is implemented by HyperSurface manufacturers and is transparently
used by developers. In its simplest form, the compiler can be seen
as a lookup table that keeps the best switch configurations corresponding
to a set of electromagnetic interactions of interest. This table is
populated by manufacturers, using well-known heuristic and analytical
techniques in physics~\cite{haupt2007genetic,Chen.2016}.

\begin{figure}[t]
\begin{centering}
\includegraphics[width=1\columnwidth]{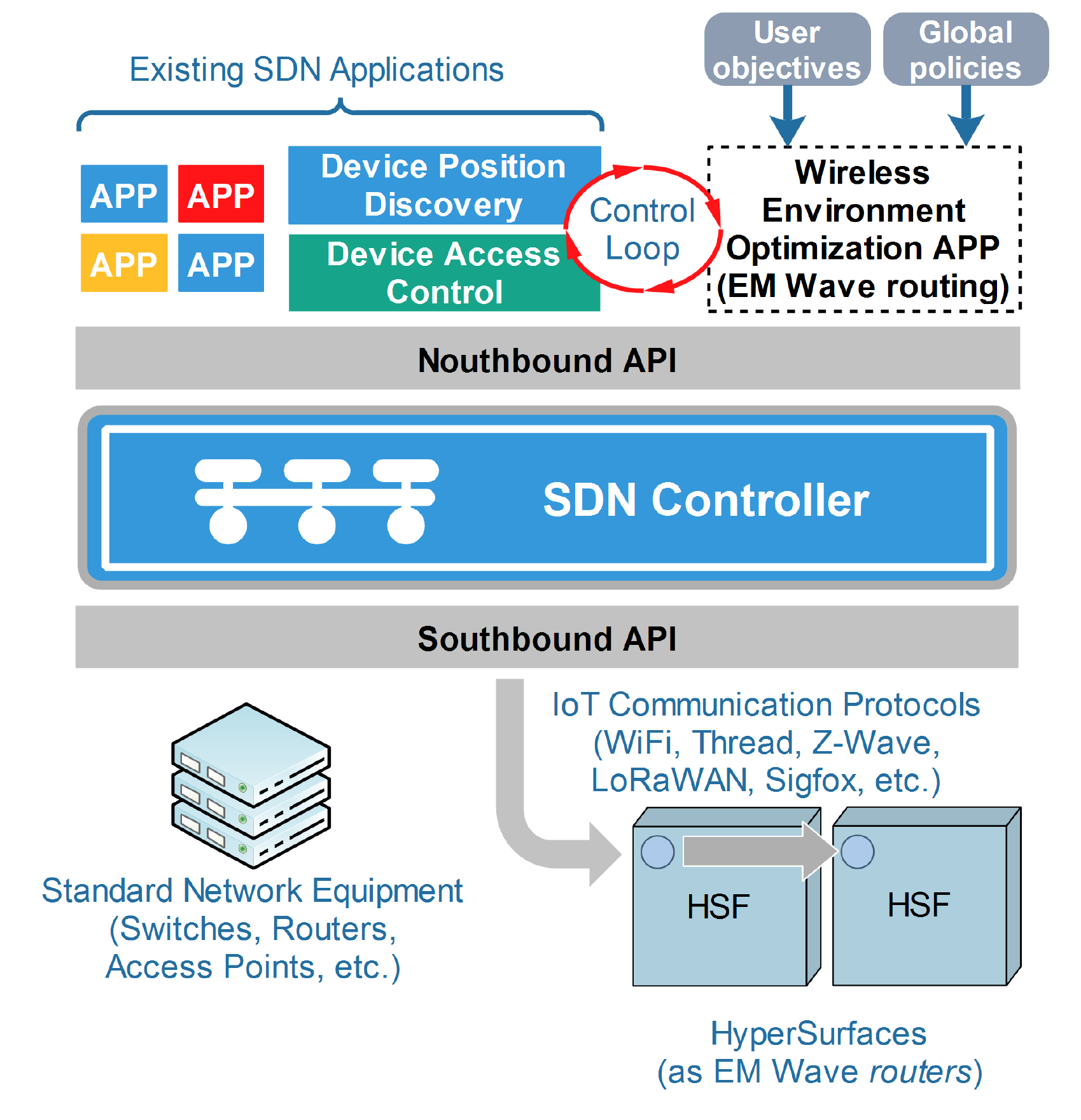}
\par\end{centering}
\caption{\label{fig:SDNarch}Integration schematic of programmable wireless
environments in the SDN paradigm.}
\end{figure}

Upon each tile there exists an IoT device that acts as its gateway.
It exerts control over the HyperSurface switches, and allows for data
exchange using common communication protocols (cf. Fig.~\ref{fig:SDNarch}).
Using these protocols, the tiles\textendash and, thus, the coated
objects\textendash become connected to common networking equipment.
Gateways of tiles upon continuous objects, such as walls, form a wired
network to facilitate power supply and the dissemination of software
commands. A selected tile acts as the object's ``representative'',
connecting to the external world.

Figure~\ref{fig:SDNarch} illustrates the integration of the programmable
wireless environment to common network infrastructure using the Software-Defined
Networking (SDN) paradigm~\cite{Akyildiz.2016}. SDN has gained significant
momentum in the past years due to the clear separation it enforces
between the network control logic and the underlying hardware. An
SDN controller abstracts the hardware specifics (``southbound''
direction) and presents a uniform programming interface (``northbound'')
that allows the modeling of network functions as applications. In
this paradigm, HyperSurface tiles are treated as wave ``routers'',
while the commands to serve a set of users, e.g., as in Fig.~\ref{fig:2dfloorplan},
are produced by a wireless environment control application. The application
takes as input the user requirements and the global policies and calculates
the fitting air paths. A control loop is established with existing
device position discovery and access control applications, constantly
adapting to environment changes.

The scalability of the novel programmable wireless environments is
a priority, both in software and hardware. In terms of software, the
additional overhead comes from the optimization service, as shown
in Fig.~\ref{fig:SDNarch}. As described, however, the optimization
pertains to finding objective-compliant paths within the graph of
tiles, which is a well-studied and tractable problem in classic networking.
In terms of hardware, the IoT gateway approach promotes miniaturization,
low manufacturing cost and minimal energy consumption of electronics,
favoring massive tile deployments to cover an environment. Moreover,
the choice of metasurfaces as the means for exerting electromagnetic
control has distinct scalability and functionality benefits over alternatives.
Metasurfaces comprise thin metallic elements and simple two-state
switches, facilitating their manufacturing using large area electronics
methods (LAE) for ultra low production cost~\cite{LAEPRINTED}. LAE
can be manufactured using conductive ink-based printing methods on
flexible and transparent polymer films, incorporating simple digital
switches such as polymer diodes~\cite{LAEPRINTED}. On the other
hand, alternatives such as antenna arrays~\cite{Moghaddam.2011}
require transceivers with accurate state control and real-time signal
processing capabilities, posing scalability considerations in terms
of size, power and manufacturing cost.

Despite their simpler design, metasurfaces constitute the state of
the art in range of wave interaction types, and with unique granularity.
Advanced frequency filtering, polarization control, and arbitrary
radiation pattern shaping functions can be potentially used for re-modulating
or ``repairing'' waves in the course of their propagation. Even
in simple wave routing and absorbing functions, metasurfaces provide
a degree of direction control so granular that has been used for the
formation of holograms~\cite{Li.2017}. A high degree of control
granularity is required for 5G, ultra-high frequency communications,
as discussed in Section~\ref{sec:Introduction}. Moreover, novel
dynamic metasurface designs employ graphene, offering operation at
the range of \emph{$\text{THz}$}~\cite{Tassin.2013}.

\section{Conclusion and ongoing efforts}

The design and implementation of HyperSurfaces is a highly interdisciplinary
task involving physics, material sciences, electrical engineering
and informatics. The combined expertise of all these disciplines results
in significant value: programmable wireless environments can be enabled
for the first time, allowing for programmatic \emph{customization}
of the laws of electromagnetic propagation, to the benefit of wireless
devices. Programmable environments provide a novel viewpoint for wireless
communications, where the usual rigid channel models are replaced
by a customizable software process. Apart from unprecedented capabilities
in wireless systems, this new view can pave the way for a completely
new class of software applications, with rich interactions with existing
security, device position discovery and user mobility prediction mechanisms
in the SDN world.

Recently, a related project under the acronym VISORSURF\footnote{A hyper\uline{visor} for meta\uline{surf}ace functions, \url{http://visorsurf.eu}.}
was funded under the prestigious Future Emerging Technologies call
of the European Union Horizon 2020 framework. VISORSURF underwent
a highly selective review phase, with just $\text{3\%}$ acceptance
rate, and attracted a total budget of $\text{5.7}$ million euros.
The multi-disciplinary team of committed researchers are actively
developing the hardware as well as the software for the HyperSurfaces
which can be applied on walls, furniture, roofs, polls and other indoor
or outdoor objects. They expect to have the first prototype within
1 $\text{\ensuremath{\nicefrac{1}{2}}}$ years, and enter mass production
soon afterwards.

\end{document}